\documentclass[preprint,5p,twocolumn]{elsarticle}

\usepackage{dcolumn}
\usepackage{bm}
\usepackage{ifpdf}
\usepackage{hyperref}
\usepackage{dcolumn}
\usepackage{bm}
\usepackage[spanish,english]{babel}
\usepackage{amsfonts}
\usepackage{amssymb}
\usepackage{graphicx}
\usepackage[latin1]{inputenc}
\usepackage{natbib}
\usepackage{geometry}
\usepackage{fleqn}

\newcommand \be{\begin{equation}}
\newcommand \en{\end{equation}}
\newcommand \bea{\begin{eqnarray}}
\newcommand \ena{\end{eqnarray}}

\journal{Physics Letters B}

\begin{document}

\begin{frontmatter}


\title{Nonsingular electrovacuum solutions with dynamically generated cosmological constant}

\author[N]{E. I. Guendelman} \ead{guendel@bgumail.bgu.ac.il}

\author[V]{Gonzalo J. Olmo}\ead{gonzalo.olmo@csic.es}

\author[J]{D. Rubiera-Garcia\corref{cor}} \ead{drubiera@fisica.ufpb.br}
\cortext[cor]{Corresponding author. Tel: +34 620775849.}

\author[N]{M. Vasihoun} \ead{maharyw@gmail.com}

\address[N]{Physics Department, Ben Gurion University of the Negev, Beer Sheva 84105, Israel}
\address[V]{Departamento de F\'{i}sica Te\'{o}rica and IFIC, Centro Mixto Universidad de
Valencia - CSIC. Universidad de Valencia, Burjassot-46100, Valencia, Spain}
\address[J] {Departamento de F\'isica, Universidade Federal da
Para\'\i ba, 58051-900 Jo\~ao Pessoa, Para\'\i ba, Brazil}

\begin{abstract}
We consider static spherically symmetric configurations in a Palatini extension of General Relativity including $R^2$ and Ricci-squared terms, which is known to replace the central singularity by a wormhole in the electrovacuum case. We modify the matter sector of the theory by adding to the usual Maxwell term a nonlinear electromagnetic extension which is known to implement a confinement mechanism in flat space. One feature of the resulting theory is that the non-linear electric field leads to a dynamically generated cosmological constant. We show that with this matter source the solutions of the model are asymptotically de Sitter and possess a wormhole topology. We discuss in some detail the conditions that guarantee  the absence of singularities and of traversable wormholes.
\end{abstract}

\begin{keyword}
Modified gravity \sep Palatini formalism \sep nonlinear electrodynamics \sep dynamical cosmological constant \sep nonsingular solutions \sep wormholes 
\end{keyword}

\end{frontmatter}

\section{Introduction}

In 1955 John Wheeler \cite{Wheeler} pointed out that well known solutions of the Einstein equations, such as Reissner-Nordstr\"om or Kerr-Newman, could be interpreted as topologically nontrivial objects connecting different regions of the space-time through a wormhole. This work, and subsequently that of Morris and Wheeler \cite{MW}, has led to several physically interesting suggestions, such as the mass-without-mass and charge-without-charge mechanisms. In this view, the electromagnetic field is not originated by a point-like charge, but instead it arises as a consequence of a flux crossing the wormhole mouth, creating the illusion of a negatively charged object on one side, and a positively charged object on the other, even though no real sources generate the field. The mass of this object would correspond to the energy stored in the electric field.

After the developments on traversable wormholes of Morris and Thorne \cite{MT}, a great deal of attention has been payed to construct wormholes from a variety of energy-momentum sources (see \cite{Visser} for a review). The standard approach consists on proposing a physically interesting wormhole metric and then drive the Einstein equations back to find the matter source that generates that geometry. More recently, motivated by the observational evidence in favour of an accelerated expansion of the Universe,  wormhole solutions including a cosmological constant have experienced renewed interest \cite{WH-cosmological} (see also the review \cite{Lemos} and references therein).

In a number of previous works \cite{or12a, lor13,or13b} the electrovacuum solutions of a simple Palatini extension of General Relativity (GR) containing $R^2$ and Ricci-squared terms in four space-time dimensions have been studied in detail by some of us. It has been found that the internal point-like singularity that typically  arises in the Reissner-Nordstrom solution of GR is generically replaced by a region of finite non-zero area that represents the mouth of a wormhole. The existence of this wormhole is an effect of the modified gravitational dynamics since the electromagnetic field does not violate any of the classical energy conditions.
The higher-order curvature terms characterizing the gravitational sector of this theory usually arise in approaches to quantum gravity \cite{string} and in the quantization of fields in curved space-time \cite{quantization}. The novelty of our approach lies on the fact that we relax the Levi-Civita condition on the metric and  obtain the field equations following the Palatini approach, in which metric and connection are regarded as two physically independent entities (see \cite{Zanelli} for a pedagogical discussion of these concepts). This implies that both metric and connection must be determined by solving their respective equations obtained through the application of the variational principle on the action. In this context one finds ghost-free, second-order field equations with Minkowski space-time as a stable vacuum solution \cite{olmo}. These properties of the quadratic Palatini theory arise because of the existence of invariant volumes associated with symmetric connections \cite{Olmo:2013lta}, and are in sharp contrast with the usual metric formulation of quadratic gravity, in which the connection is imposed to be the Levi-Civita one \emph{a priori}, resulting in fourth-order field equations generically affected by ghosts. It is worth noting that while in the metric formalism there exists a  family of Lagrangians leading to second-order equations (Lovelock gravities \cite{Lovelock}), the extra terms become topological invariants in a four-dimensional space-time, and the theory provides the same dynamics as GR. In contrast, the Palatini approach yields nontrivial modified dynamics in four dimensions as long as matter fields are present because they play an active role in the determination of the connection. In vacuum, our theory boils down to GR, which is a manifestation of the universality of Einstein's equations observed in Palatini theories \cite{Francaviglia} (see also \cite{Olmo:2013lta}).

In this work we present wormhole-type solutions with a dynamically generated cosmological constant in the quadratic Palatini extension of GR studied in \cite{or12a, lor13,or13b}. The matter sector of our theory, responsible for the dynamical generation of a cosmological constant \cite{Guendelmana} (see also \cite{Habib}), will be described by a nonlinear theory of electrodynamics given by the Lagrangian density
\begin{equation} \label{lagrangian-confining}
\varphi(X)=X -g \sqrt{2X},
\end{equation}
where $X=-\frac{1}{2} F_{\mu\nu}F^{\mu\nu}$ and $F_{\mu\nu}=\partial_{\mu}A_{\nu}- \partial_{\nu}A_{\mu}$ is the field strength tensor of the vector potential $A_{\mu}$. In the absence of gravity, the square root term in (\ref{lagrangian-confining}) naturally arises as a spontaneous breakdown of the scale symmetry of the Maxwell lagrangian $X$ \cite{Guendelman-0}, being $g>0$ an integration constant responsible for this breakdown.
Moreover, when coupled to charged fermions, the model (\ref{lagrangian-confining}) produces a confinement effective potential $V(r)=-\frac{q}{r}+\frac{g}{\sqrt{2}}r$, which is of the form of the well known Cornell potential, which has been used for the effective description of heavy quark-antiquark strong interactions \cite{Cornell}. This implements 't Hooft's description of linear confinement phenomena \cite{tHooft} since the electromagnetic field energy becomes a linear function of the electric displacement field in the infrared region. Let us note that one could start with the non-abelian version of (\ref{lagrangian-confining}) and for static spherically symmetric solutions the nonabelian theory effectively boils down to the abelian one, as pointed out in \cite{Guendelman-0} (see also \cite{dr08}). Here we shall see that the coupling of this field to a quadratic Palatini model yields a dynamically generated cosmological constant for large distances and, in addition, the GR singularity is generically replaced by a wormhole.  This work complements the results recently found in \cite{or12a,lor13}, extending them to the case of nonlinear electrodynamics with modifications in the infrared sector.

\section{General formalism}

We consider a family of Palatini theories defined as
\begin{equation}\label{eq:action}
S[g,\Gamma,\psi_m]=\frac{1}{2\kappa^2}\int d^4x \sqrt{-g}f(R,Q) +S_m[g,\psi_m],
\end{equation}
where $f(R,Q)$ represents the gravity Lagrangian, $\kappa^2$ is a constant with suitable dimensions (in GR, $\kappa^2 \equiv 8\pi G$),  $\Gamma \equiv \Gamma_{\mu\nu}^{\alpha}$ is the independent connection, $g_{\alpha\beta}$ is the space-time metric,  $R=g^{\mu\nu}R_{\mu\nu}$, $Q=g^{\mu\alpha}g^{\nu\beta}R_{\mu\nu}R_{\alpha\beta}$, $R_{\mu\nu}={R^\rho}_{\mu\rho\nu}$, ${R^\alpha}_{\beta\mu\nu}=\partial_{\mu}
\Gamma^{\alpha}_{\nu\beta}-\partial_{\nu}
\Gamma^{\alpha}_{\mu\beta}+\Gamma^{\alpha}_{\mu\lambda}\Gamma^{\lambda}_{\nu\beta}-
\Gamma^{\alpha}_{\nu\lambda}\Gamma^{\lambda}_{\mu\beta} $ is the Riemann tensor, and $S_m$ the matter action. For concreteness, in this work we shall focus on the quadratic Lagrangian
\begin{equation} \label{eq:gravitytheory}
f(R,Q)=R + l_P^2(a R^2+ b Q) \ ,
\end{equation}
where $l_P \equiv \sqrt{\hbar G/c^3}$ is the Planck length, and $a$ and $b$ are dimensionless parameters. Performing independent variations of the action (\ref{eq:action}) with respect to metric and connection leads to
\begin{eqnarray}
f_R R_{\mu\nu}-\frac{f}{2}g_{\mu\nu}+2f_QR_{\mu\alpha}{R^\alpha}_\nu &=& \kappa^2 T_{\mu\nu}\label{eq:met-varX}\\
\nabla_{\beta}\left[\sqrt{-g}\left(f_R g^{\mu\nu}+2f_Q R^{\mu\nu}\right)\right]&=&0  \ ,
 \label{eq:con-varX}
\end{eqnarray}
where $f_R \equiv \frac{df}{dR}$, $f_Q \equiv \frac{df}{dQ}$, and $T_{\mu\nu}$ is the energy-momentum tensor of the matter. For simplicity, we have set the torsion to zero, which guarantees that $R_{[\mu\nu]}=0$ (see \cite{Olmo:2013lta} for details). To solve Eqs.(\ref{eq:met-varX}) and (\ref{eq:con-varX}) we introduce the matrix  $\hat{P}$ (whose components are  ${P_\mu}^\nu\equiv R_{\mu\alpha}g^{\alpha\nu}$), which allows us to express (\ref{eq:met-varX}) as
\begin{equation}
2f_Q\hat{P}^2+f_R \hat{P}-\frac{f}{2}\hat{I} = \kappa^2 \hat{T} \label{eq:met-varRQ2} \ ,
\end{equation}
where $\hat{T}$ is the matrix representation of ${T_\mu}^\nu$. Since in this notation we have $R={[\hat{P}]_\mu}^\mu$ and $Q={[\hat{P}^2]_\mu}^\mu$, we can see  (\ref{eq:met-varRQ2}) as a nonlinear algebraic equation for  $\hat{P}=\hat{P}(\hat{T})$. Bearing  in mind the relation $\hat{P}=\hat{P}(\hat{T})$, the connection equation (\ref{eq:con-varX}) can be written as
\begin{equation} \label{eq:auxmetric}
\nabla_{\beta}[\sqrt{-g} g^{\mu\alpha} {\Sigma_\alpha}^\nu]=0 \ ,
\end{equation}
 where we have introduced the object
\begin{equation} \label{eq:geometrymissmatch}
{\Sigma_\alpha}^{\nu}=\left(f_R \delta_{\alpha}^{\nu} +2f_Q {P_\alpha}^{\nu}\right) \ .
\end{equation}
Note that since ${\Sigma_\alpha}^\nu$ and $g^{\mu\alpha}$ do not depend explicitly on $\Gamma_{\mu\nu}^{\alpha}$, the connection in (\ref{eq:con-varX}) appears linearly and can be solved by algebraic means. This motivates the introduction of a symmetric rank-two tensor $h^{\mu\nu}$ satisfying
\begin{equation} \label{eq:auxmetric}
\nabla_{\beta}[\sqrt{-g} g^{\mu\alpha} {\Sigma_\alpha}^\nu]=\nabla_{\beta}[\sqrt{-h} h^{\mu\nu}]=0 \ .
\end{equation}
The existence of $h_{\mu\nu}$ implies that $\Gamma^\alpha_{\mu\nu}$ is the Levi-Civita connection of $h_{\mu\nu}$  \cite{Olmo:2013lta}. Comparison of the terms within brackets in this equation leads to the following solution
{
\begin{equation} \label{eq:h-g}
h^{\mu\nu}=\frac{g^{\mu\alpha}{\Sigma_{\alpha}}^\nu}{\sqrt{\det \hat{\Sigma}}} \ , \quad
h_{\mu\nu}=\left(\sqrt{\det \hat{\Sigma}}\right){\Sigma_{\mu}}^{\alpha}g_{\alpha\nu} \ .
\end{equation}
Using the definition of ${\Sigma_\mu}^\nu$ and the relations (\ref{eq:h-g}), it is easy to see that (\ref{eq:met-varX})  can be written as ${P_\mu}^\alpha {\Sigma_\alpha}^\nu=R_{\mu\alpha}h^{\alpha\nu} \sqrt{\det \hat\Sigma}=\frac{f}{2}{\delta_\mu^\nu}+{T_\mu}^\nu$, which allows to express the metric field equations using $h_{\mu\nu}$ in the compact form
\begin{equation} \label{eq:fieldequations}
{R_{\mu}}^{\nu}(h)=\frac{1}{\sqrt{\det \hat{\Sigma}}}\left(\frac{f}{2}{\delta_{\mu}}^{\nu}+ \kappa^2 {T_{\mu}}^{\nu} \right) \ .
\end{equation}

\section{The equations and solutions for nonlinear electrodynamics}

For the sake of generality, we shall consider as the matter action in (\ref{eq:action}) nonlinear electrodynamics defined as
\begin{equation} \label{eq:NEDaction}
S_m=\frac{1}{8\pi} \int d^4x \sqrt{-g} \varphi(X,Y),
\end{equation}
where $\varphi(X,Y)$ is a given function of the two field invariants $X=-\frac{1}{2}F_{\mu\nu}F^{\mu\nu}$ and $Y=-\frac{1}{2}F_{\mu\nu}{*F}^{\mu\nu}$, where $*F^{\mu\nu}=\frac{1}{2}\epsilon^{\mu\nu\alpha\beta}F_{\alpha\beta}$ is the dual of $F_{\mu\nu}$. The matter field equations,
$\nabla_{\mu} \left( \sqrt{-g} (\varphi_X F^{\mu\nu} + \varphi_Y {*F}^{\mu\nu} \right))=0$,
admit, for purely electrostatic configurations, $E(r)=F^{tr}(r)$, and assuming a line element of the form $ds^2=g_{tt}dt^2+g_{rr}dr^2+r^2d\Omega^2$, a first integral of the form
\begin{equation} \label{eq:first-integral}
F^{tr}=\frac{q}{r^2 \varphi_X \sqrt{-g_{tt}g_{rr}}},
\end{equation}
where $q$ is an integration constant. Writing $X=-g_{tt}g_{rr} (F^{tr})^2$ (note that for these solutions $Y=0$) it follows that for any spherically symmetric metric
\begin{equation}
\varphi_X^2 X=\frac{q^2}{r^4}.
\end{equation}
The energy-momentum tensor obtained from (\ref{eq:NEDaction})  reads
\begin{equation} \label{eq:em1}
{T_{\mu}}^{\nu}=-\frac{1}{4\pi}\left[\varphi_X {F_{\mu}}^{\alpha}{F_{\alpha}}^{\nu}+ \varphi_Y {F_{\mu}}^{\alpha}{*F_{\alpha}}^{\nu}-\frac{\delta_{\mu}^{\nu}}{4}\varphi\right],
\end{equation}
which means that we can write its components as
\begin{equation}\label{eq:Tmn-EM}
{T_\mu}^\nu=\frac{1}{8\pi} \left( \begin{array}{cc}
 (\varphi-2X\varphi_X) \hat{I} & \hat{0}  \\
\hat{0} & \varphi \hat{I} \\
\end{array} \right),
\end{equation}
where $\hat{I}$ and $\hat 0$ are the identity and zero $2 \times 2$ matrices, respectively. Next, to find the explicit form of $\hat{P}$ for our problem (necessary to obtain $\hat{\Sigma}$) we write (\ref{eq:met-varRQ2}) as
\begin{equation}
2f_Q\left(\hat{P}+\frac{f_R}{4f_Q}\hat{I}\right)^2= \left(
\begin{array}{cc}
\lambda_-^2\hat{I} & \hat{0}  \\
\hat{0} & \lambda_+^2\hat{I} \\
\end{array} \right),  \label{eq:P}
\end{equation}
where $ \lambda_{+}^2=\frac{1}{2}\left(f+\frac{f_R^2}{4f_Q}+2k^2 T_{\theta}^{\theta}\right) \label{lambda+},$ and $\lambda_{-}^2=\frac{1}{2} \left(f+\frac{f_R^2}{4f_Q} +2k^2 T_{t}^{t}\right) $. Taking the square root in (\ref{eq:P}) and demanding agreement with GR in the low curvature regime (where $f_R\to 1$ and $f_Q\to 0$) we obtain
\begin{equation}\label{eq:M_ab}
\sqrt{2f_Q}\left(\hat{P}+\frac{f_R}{4f_Q}\hat{I}\right)= \left(
\begin{array}{cc}
\lambda_- \hat{I}& \hat{0} \\
\hat{0} & \lambda_+ \hat{I} \\
\end{array}
\right).
\end{equation}
From this it follows that the matrix $\hat{\Sigma}$ is given by
\begin{equation} \label{eq:sigma-matrix}
\hat{\Sigma}=\frac{f_R}{2}\hat{I}+\sqrt{2f_Q} \left(
\begin{array}{cc}
\lambda_- \hat{I}& \hat{0} \\
\hat{0} & \lambda_+ \hat{I} \\
\end{array} \right)= \left(
\begin{array}{cc}
\sigma_- \hat{I}& \hat{0} \\
\hat{0} & \sigma_+\hat{I} \\
\end{array}
\right),
\end{equation}
where $\sigma_\pm=\left(\frac{f_R}{2}+\sqrt{2f_Q}\lambda_\pm\right)$. Gathering all these elements we obtain the field equations
for our model
\begin{equation}\label{eq:Rmn}
{R_{\mu}}^{\nu}(h)=\frac{1}{2 \sigma_{+} \sigma_{-}} \left(
\begin{array}{cc}
(f+2\kappa^2 T_t^t)\hat{I} & \hat{0} \\
\hat{0} & (f+2\kappa^2 T_{\theta}^{\theta}) \hat{I} \\
\end{array}
\right).
\end{equation}
Focusing now on the gravity Lagrangian (\ref{eq:gravitytheory}), tracing in Eq.(\ref{eq:met-varX}) with $g_{\mu\nu}$ yields $R=-\kappa^2 T$ where $T \neq 0$ is the trace of the energy-momentum tensor.  On the other hand, taking the trace in (\ref{eq:M_ab}) we obtain $Q$ as
\be \label{eq:Q-invariant}
Q= \kappa^4 \left( \frac{T^2}{4}+\frac{(T_t^t- T_{\theta}^{\theta})^2}{(1-(2a+b) \kappa^2 l_P^2 T)^2} \right).
\en
Major simplifications arise when $b=1$ and $a=-1/2$ and, therefore, from now on we will stick ourselves to that particular case, which leads to
\begin{eqnarray} \label{eq:sigma1}
\sigma_{+}&=&1-2\kappa^2l_P^2 T_{\theta}^{\theta}\\
\sigma_{-}&=& 1-2\kappa^2l_P^2 T_{t}^{t}. \label{eq:sigma2}
\end{eqnarray}
To solve the field equations (\ref{eq:Rmn}) we introduce two different line elements in Schwarzschild-like coordinates, one associated to the physical metric $g_{\mu\nu}$
\begin{equation}\label{eq:ds2g}
ds^2= g_{tt}dt^2+g_{rr}dr^2+r^2 d\Omega^2 \ ,
\end{equation}
which has been used in solving for the electrostatic field,
and another associated to the auxiliary metric $h_{\mu\nu}$
\begin{equation}\label{eq:ds2h}
d\tilde{s}^2= h_{tt}dt^2+h_{rr}dr^2+\tilde{r}^2 d\Omega^2,
\end{equation}
with $d\Omega^2= d\theta^2+\sin^2 (\theta) d \phi^2$. The relation between these two line elements is obtained via $g_{\mu\nu}={\Sigma_\mu}^{\alpha}h_{\alpha\nu}/\sqrt{\det \Sigma}$, which implies that $g_{tt}=h_{tt}/\sigma_+$, $g_{rr}=h_{rr}/\sigma_+$, and $\tilde{r}^2 =r^2 \sigma_{-}$. Next, we find it useful to  use $\tilde{r}$ as the radial coordinate, which brings  (\ref{eq:ds2h}) into
\begin{equation}\label{eq:ds2h}
d\tilde{s}^2= h_{tt}dt^2+h_{\tilde{r}\tilde{r}}d\tilde{r}^2+\tilde{r}^2 d\Omega^2 \ ,
\end{equation}
where $h_{rr}dr^2=h_{\tilde{r}\tilde{r}}d\tilde{r}^2$. Using the ansatzes
$h_{tt}=-A(\tilde{r})e^{2\psi(\tilde{r})}, h_{\tilde{r}\tilde{r}}=1/A(\tilde{r})$, 
one finds that $\psi=0$, and that $A(\tilde{r})$ satisfies
\begin{equation} \label{eq:eomunique}
\frac{1}{\tilde{r}^2} \left(1-A(\tilde{r})-\tilde{r}A_{\tilde{r}} \right)=\frac{1}{2\sigma_+ \sigma_-}\left(f+\frac{\kappa^2}{4\pi} \varphi \right) \ .
\end{equation}
Taking the ansatz $ A(\tilde{r})=1-\frac{2M(\tilde{r})}{\tilde{r}}$ and using the relation between coordinates $\tilde{r}^2=r^2 \sigma_{-}$, which implies  $\frac{d\tilde{r}}{dr}=\sigma_{-}^{1/2}\left(1+\frac{r\sigma_{-,r}}{2\sigma_{-}} \right)$, we can write Eq.(\ref{eq:eomunique}) as
\begin{equation} \label{eq:mass-r}
\frac{dM}{dr}=\frac{\left(f+\frac{\kappa^2}{4\pi}\varphi\right)r^2 \sigma_{-}^{1/2}}{4\sigma_+}\left(1+\frac{r\sigma_{-,r}}{2\sigma_-}\right).
\end{equation}
Upon integration of this expression we would obtain an expression of the form $ \frac{M(r)}{M_0}=1+\delta_1 G(r)$, where $M_0$ is an integration constant identified as the Schwarzschild mass, $M_0\equiv r_S/2$, and we have isolated the constant
\begin{equation}\label{eq:d1d2}
\delta_1=\frac{1}{2r_S}\sqrt{\frac{r_q^3}{l_P}} .
\end{equation}
The function $G(r)$  satisfies
\begin{equation} \label{eq:Gzz}
\frac{dG}{dr}=\frac{r^2 \sigma_{-}^{1/2}}{\sigma_+}\left(1+\frac{r\sigma_{-,r}}{2\sigma_-}\right)\left(\frac{4\pi}{\kappa^2}f+\varphi\right) \ .
\end{equation}
With all these elements we arrive to the final expression for the metric components in (\ref{eq:ds2g}) as
\begin{eqnarray}
g_{tt}=-\frac{A(r)}{\sigma_{+}(r)} &;& g_{rr}=\frac{\sigma_{-}(r) }{\sigma_{+}(r)A(r)} \left(1+\frac{r\sigma_{-,r}}{2\sigma_{-}(r)} \right)^2\label{eq:A1}\\
A(r)&=&1-\frac{1+\delta_1 G(r)}{\delta_2 r \sigma_{-}(r)^{1/2}} \label{eq:A2} \ ,
\end{eqnarray}
where we have defined $\delta_2= \frac{\sqrt{r_q l_P}}{r_S}$.
Eqs.(\ref{eq:A1}) and (\ref{eq:A2}) together with (\ref{eq:Gzz}) and $\sigma_{\pm}$ fully characterize our problem, once the function $\varphi(X)$ is specified.

\section{Confining electromagnetic field}

From the Lagrangian density (\ref{lagrangian-confining}), using the field equations (\ref{eq:first-integral}), the $X$-invariant takes the form
\begin{equation} \label{field-confining}
X=\left(\frac{q}{r^2}+\frac{g}{\sqrt{2}} \right)^2,
\end{equation}
while the energy-momentum tensor components are
\begin{eqnarray}
T_t^t&=&-\frac{X}{8\pi}=-\frac{1}{8\pi}\left(\frac{q^2}{r^4}+\frac{\sqrt{2}g q}{r^2}+\frac{g^2}{2} \right) \\ T_{\theta}^{\theta}&=&\frac{1}{8\pi} \left(X-\sqrt{2}g X^{1/2} \right)=\frac{1}{8\pi} \left(\frac{q^2}{r^4}-\frac{g^2}{2} \right).
\end{eqnarray}
Defining $r_q^2 \equiv \widetilde{\kappa}^2 q^2$ and $r_g^2\equiv (\widetilde{\kappa}^2 g^2/2 )^{-1}$,  introducing the dimensionless variable  $z=r/r_c$, with $r_c\equiv \sqrt{l_P r_q}$, and denoting $\delta_3\equiv l_P/r_g$,  Eqs.(\ref{eq:sigma1}) and (\ref{eq:sigma2}) lead to
\begin{equation} \label{eq:sigma-confining}
\sigma_{+}=1+\frac{1}{z^4} - \delta_3^2 \hspace{0.2cm} ; \hspace{0.2cm} \sigma_{-}=1 - \frac{1}{z^4} -\frac{2 \delta_3}{z^2} - \delta_3^2.
\end{equation}
With the same notation, the function $G_z$ in (\ref{eq:Gzz}) becomes
\begin{equation} \label{eq:Gz}
\frac{dG}{dz}=\left(1+\delta_3 z^2 \right)^2\left(\frac{z^4(1-\delta_3^2)+1}{z^4 \sqrt{z^4(1-\delta_3^2) - 2 \delta_3 z^2 -1}}\right).
\end{equation}
This choice of variables highlights the three fundamental scales present in the problem, namely, the charge-to-mass ratio $r_q/r_S$, the charge-to-Planck ratio $r_q/l_P$ and the Planck-to-nonlinear electrodynamics ratio $l_P/r_g$. Note that when $\delta_3 \rightarrow 0$ (so $g=0$ in (\ref{field-confining})) we have $\frac{dG}{dz}=\frac{z^4+1}{z^4 \sqrt{z^4-1}}$, which recovers the Palatini $f(R,Q)$-Maxwell solutions considered in Ref.\cite{or12a}.

To integrate $G_z$ we note that this function is only defined for $z>z_c=\frac{1}{\sqrt{1-\delta_3}}$, for which the denominator of (\ref{eq:Gz}) vanishes. This corresponds to the point where the 
 function $\sigma_{-}$ vanishes and is intimately related to the existence of a wormhole. In a sense, the $z=z_c$ surface defines a core that replaces the usual point-like singularity of GR. Taking into account the existence of this core we can exactly integrate the function $G_z$. To do this we find it useful to introduce the following change of variable
\begin{equation}\label{eq:z2x}
z=\frac{x}{\sqrt{1-\delta_3}}=x\sqrt{\frac{1+\chi}{{2}}} \ ,
\end{equation}
where we have introduced the constant $\chi$ through $\delta_3=\frac{\chi-1}{\chi+1}$, with $\chi\ge 1$ and such that $\chi \rightarrow 1$ implies $\delta_3 \rightarrow 0$ (the Maxwell limit). With these definitions the function $G_x=G_z dz/dx$ becomes
\begin{equation}\label{eq:Gx3}
G_x= \frac{\left((\chi -1) x^2+2\right)^2 \left(\chi  x^4+1\right)}{\sqrt{2} (\chi +1)^{3/2} x^4 \sqrt{\left(x^2-1\right) \left(\chi  x^2+1\right)}}.
\end{equation}
The integration of $G_x$ is analytical, and gives
\begin{eqnarray}
G(x)&=& \delta(\chi)+ \frac{\sqrt{\left(x^2-1\right)  \left(\chi  x^2+1\right)} \left((\chi -1) x^2+2\right)^2}{3 \sqrt{2}  (\chi +1)^{3/2} x^3} \nonumber \\
&+&\frac{i\sqrt{2(\chi +1)}}{3 \chi } \Big[(\chi+1) {_E F}\left(\arcsin (x),-\chi \right) +\nonumber \\
&+& (\chi-1) {_E E}\left(\arcsin(x),-\chi \right) \Big],   \label{eq:Gx}
\end{eqnarray}
where ${_E F}$ and ${_E E}$ are the elliptic integrals of the second and first kind, respectively, and $\delta(\chi)=\beta(\chi)-i \gamma(\chi)$ is a  constant needed to keep $G(x)$ real and to recover the GR value when $x\gg 1$. Its value is
\begin{eqnarray}
\beta(\chi)&=& \frac{\sqrt{2(\chi+1)}}{3\chi^{3/2}} \Big[ (\chi-1) \Big(\chi \  {_E E}(-\chi^{-1}) \nonumber \\ &-& (\chi+1) \ {_E K}(-\chi^{-1}) \Big)   \\
&-& \sqrt{\chi} (\chi+1)( {_E K}(1+\chi) + i{_E K}(-\chi) ) \Big] \nonumber \\
\gamma(\chi)&=& \frac{\sqrt{2(\chi+1)}}{3\chi}\Big[(\chi-1){_E E}(-\chi) \nonumber \\ &+& (\chi+1){_E  K} (-\chi)\Big] ,
\end{eqnarray}
where ${_E E}$ and ${_E K}$ are complete elliptic integrals of the first kind. 
 We note that when $\chi \rightarrow 1$ ($\delta_3 \rightarrow 0$) we obtain $\beta(1)  \simeq -1.74804$, which recovers the value corresponding to the Maxwell case \cite{or12a}.

\subsection{Geometry and topology as $x\to 1$}

To study the behaviour of the metric near the core, $x\to 1$, we find it useful to write the line element (\ref{eq:ds2g}) as
\begin{equation}
ds^2=g_{tt}dt^2-g_{tt}^{-1} r_c^2 dz^{*2} +r_c^2 z^2(z^*)d\Omega^2,
\end{equation}
where the new coordinate $z^*$ satisfies $(dz^*/dz)^2=1/\sigma_{-}$. This representation puts forward that $g_{tt}(z^*)$ and $z^2(z^*)$ contain all the information on the geometry. An $x^*$ coordinate can also be defined in analogy with (\ref{eq:z2x}) as $z^*={x^*}/{\sqrt{1-\delta_3}}$. The relation between $x$ and $x^*$ is found by direct integration as $x^*(x)=W(x,\chi)$, where
\begin{equation}\label{eq:Wxchi}
W(x,\chi)=w(\chi) - \frac{(\chi+1)^{3/2}}{2 \sqrt{2} \chi} \left[F_1(x,\chi)-F_2(x,\chi)\right] \ ,
\end{equation}
and we have defined $F_1(x,\chi)\equiv {_E{F}}\left(\arcsin (x),-\chi \right)$, $F_2(x,\chi)\equiv {_E{E}}\left(\arcsin (x),-\chi \right)$, and the constant
\begin{eqnarray}
w(\chi) &=& \frac{(\chi+1)^{3/2}}{2\sqrt{2} \chi^{3/2}} \Big[\chi^{1/2} \Big({_E{K}}(\chi+1) + i {_E{K}}(-\chi) \Big) \nonumber \\
&-&  (\chi +1) {_E{K}}\left(-\chi^{-1}\right) +\chi \  {_E{E}}\left(-\chi^{-1} \right)+\nonumber \\ &+&i\chi^{1/2} \Big( {_E{K}}(-\chi )- {_E{E}}(-\chi )\Big) \Big]  \ .
\end{eqnarray}
As $x \rightarrow 1$ we obtain $x^* \simeq w(\chi)+ \frac{1}{2} (\chi +1) \sqrt{x-1} + \ldots$.
From the explicit expression of $G(x)$ obtained in (\ref{eq:Gx}), $\sigma_{\pm}$ in Eq.(\ref{eq:sigma-confining}) and the general formulae (\ref{eq:A1}) and (\ref{eq:A2}) we can perform series expansions of $g_{tt}$ around $x=1$ as
\begin{equation}\label{eq:gtt_series}
g_{tt} \simeq  \frac{(\chi +1) ({\delta_c^\chi}-\delta_1)}{8 \delta_2 {\delta_c^\chi} \sqrt{x-1}} + \frac{(\chi +1) (\delta_1 (\chi +1)-2 \delta_2)}{8 \delta_2}+\ldots
\end{equation}
where ${\delta_c^\chi}=-1/ \beta(\chi)$. While this function is in general divergent as $x \rightarrow 1$, for $\delta_1={\delta_c^\chi}$ we find that (\ref{eq:gtt_series}) is finite.
Note, in this sense, that through a constant rescaling of $t$ and $x^{*}$ the line element near $x=1$ describes a Minkowskian-like region. As a result, curvature invariants such as $R(g)$, $Q(g)$, and the Kretschmann scalar
become finite at $x=1$ for arbitrary $\chi$ when $\delta_1={\delta_c^\chi}$. In fact,
$r_c^4K(g)|_{x=1}=r_c^4 R_{\alpha \beta \gamma \delta}(g)R^{\alpha \beta \gamma \delta}(g)_{x=1} \simeq  4(4 + {2[{\delta_c^\chi}(\chi+1)/\delta_2-2]^2} + {[{\delta_c^\chi}(\chi+1)(\chi+3)/\delta_2-12]^2}/{9 (\chi + 1)^2})/(\chi+1)^2$.
We note that for $\delta_1 \neq {\delta_c^\chi}$ those scalars diverge at  $x=1$.  However, the existence of the solutions $\delta_1={\delta_c^\chi}$, for which the geometry at $x=1$ is smooth, drives us to consider an extension of the geometry beyond that point. In this sense, note that in the definition of $x^*$ we assumed that $dx^*=dx/\sigma_{-}^{1/2}$ and discarded the possibility of having $dx^*=-dx/\sigma_{-}^{1/2}$. Taking this into account, an extension covering the whole range $-\infty<x^*<\infty$ is possible if $x^*(x)$ is redefined using (\ref{eq:Wxchi}) as
\begin{equation} \label{eq:wormhole}
x^*(x)=\left\{\begin{array}{lr} W(x,\chi) & if \hspace{0.2cm} x^*\ge w(\chi) \\
                                              2 w(\chi) -W(x,\chi) & if  \hspace{0.2cm} x^*\le w(\chi)\end{array}\right.
\end{equation}
Thus the divergence of $dx^{*}/dx$ at $x=1$ simply states that the coordinate $x(x^*)$ has reached a minimum there. The branch with $dx^{*}/dx<0$ describes a new region accross $x=1$ in which the area of the $2$-spheres grows as $x^{*} \rightarrow - \infty$, and gives continuity to the function $G(x)$  accross the {\it bounce} at $x^*=w(\chi) $.  Thus, to cover the whole geometry in terms of the coordinates $(t,z)$, or $(t,x)$, one needs to use two charts, one for the region $x^* \geq w(\chi) $ and another for $x^* \leq w(\chi) $.  This means that our solution has a genuine wormhole structure connecting two regions of space-time through a spherical tunnel of area $A_c=4\pi r_c^2 z_c^2=4\pi r_c^2 /(1-\delta_3)$.

It is worth noting that an observer in the $x^*>w(\chi)$ (or $x^*<w(\chi)$) region would measure an electric flux through any $2$-surface $S$ enclosing the wormhole throat of magnitude $\Phi \equiv \int_S \varphi_X =4\pi q$ (or $-4\pi q$ if $x^*<w(\chi)$, due to the change of orientation in the normal to $S$). Moreover, the flux per surface unit turns out to be $\frac{\Phi }{4\pi r_c^2 z_c^2}= (1-\delta_3) \sqrt{\frac{c^7}{2(\hbar G)^3} }$ which is independent of the specific amounts of mass and charge or, equivalently, of the specific values of $\delta_1$ and $\delta_2$. This fact puts forward that the (topological) wormhole structure is also present for those solutions with $\delta_1\neq \delta_1^\chi$, which exhibit curvature divergences at the wormhole throat. Note that it is a remarkable fact that the existence of curvature divergences is not an obstacle to having a well-defined electric flux through the wormhole throat. The topologically nontrivial structure of these solutions implies that the spherically symmetric field that local observers measure is not generated by a distribution of charges, but instead is a consequence of an electric flux trapped in the topology \cite{Wheeler}, which  avoids the (traditional  and implicit) picture of source charges compressed at infinite density at the singularity.

A graphical representation reveals that the sign of the $g_{tt}$ component from Eqs.(\ref{eq:A1}), (\ref{eq:A2}), (\ref{eq:sigma-confining}) and (\ref{eq:Gx}) determines whether an event horizon exists or not. In this sense, the event horizon dissappears if ${\delta_c^\chi}/\delta_1=r_q/2l_P<(1-\delta_3)$. This becomes more transparent if we write the charge as $q=N_q e$, where $N_q$ is the number of charges and $e$ the electron charge. The horizon disappears if $N_q<N_q^c$, where $N_q^c = (1-\delta_3) \sqrt{2/\alpha_{em}} \simeq 16.55 (1-\delta_3) $ is the critical number of charges and $\alpha_{em}$ the fine structure constant.

\subsection{de Sitter behavior when $x\gg 1$}

Expanding $g_{tt}$ in $x \gg 1$, restoring the usual notation, and keeping only up to second-order terms, we find
\begin{eqnarray} \label{eq:asymp}
g_{tt} &\simeq &-\left(1-\frac{r_q}{r_g}+\frac{l_P^2}{r_g^2}\right)+\frac{r_S }{r}-\frac{r_q^2}{2 r^2} +\frac{r^2}{6 r_g^2}  \\ &+&O\left(\frac{1}{r^3}\right) \nonumber .
\end{eqnarray}
This expression is not exactly that of a Reissner-Nordstr\"{o}m-de-Sitter solution due to the presence of the terms $-\frac{r_q}{r_g}+\frac{l_P^2}{r_g^2}$, resulting from the corrections to the Maxwell Lagrangian in (\ref{lagrangian-confining}). The appareance of a cosmological constant term $r^2/6r_g^2$ in (\ref{eq:asymp}) is a similar effect to that found in \cite{Guendelmana}. In (\ref{eq:asymp}) we have kept the minimum number of terms containing both ultraviolet corrections due to the gravitational Lagrangian, which manifest the existence of a wormhole structure, and  infrared corrections coming from the matter sector, which are responsible for the asymptotically de Sitter behavior. At third and higher orders, $l_P/r_g$ terms also add corrections to $r_S$, $r_q$, and $\Lambda_{eff}\equiv 1/2r_g^2$.
 For realistic values of the parameters, the Planck scale corrections are completely irrelevant as soon as $r$ is a few units larger than $l_P$, as can be verified by explicit evaluation of the curvature invariants.

 Let us recall that no bare cosmological constant  $\Lambda_B$ was included in the starting action (\ref{eq:action}). Our  $\Lambda_{eff}$ arises dynamically because of the confining electric field (\ref{field-confining}) associated to the Lagrangian (\ref{lagrangian-confining}). The introduction of a $\Lambda_B$ in the action can be carried out efficiently by considering a matter Lagrangian of the form $\tilde{\varphi}(X)=\varphi(X)+\varphi_B$ with $\varphi_B\equiv -8\pi \Lambda_B/\kappa^2$. Its effects are non-trivial, and  (\ref{eq:Gx3}) picks up a new term (which can be solved exactly)
\begin{equation}
G_x^{\Lambda}=G_x-\delta_4^2\frac{ A(x,\chi)+\delta_4^2 B(x,\chi)}{4 \sqrt{2} \sqrt{\left(x^2-1\right) \left(\chi  x^2+1\right)}} \ ,
\end{equation}
where $A(x,\chi)=\sqrt{\chi +1} x^2 \left(4 (\chi -1)+\left(\chi ^2-6 \chi +1\right) x^2\right)$, $B(x,\chi)=(\chi +1)^{5/2} x^4$, and $\delta_4^2\equiv 2\Lambda_B l_P^2$.  The metric at $x\to 1$ behaves as in (\ref{eq:gtt_series}) but with $\delta_c^\chi$ containing $\chi$-dependent contributions proportional to $\delta_4^2$ and $\delta_4^4$, and the constant term replaced by $ (\chi +1) \left(\delta _1 \left(\delta _4^2+1\right) (\chi +1) \left(4-\delta _4^2 (\chi +1)\right)-8 \delta _2\right)/32 \delta _2$. In the far limit, we find that the leading order correction produces the shift $\Lambda_{eff}\to \Lambda_B+1/2r_g^2$. Higher order contributions also add corrections to  $r_S$, $r_q$, and $\Lambda_{eff}$.

It is worth noting that the sign of  $g$ does not alter the value of the  cosmological term $1/2r_g^2$, which is always positive. Nonetheless, this sign would be observable in regions of weak gravitational field but with a measurable ratio $r_q/r_g$, where the half-life of unstable particles, for instance, would be different from that expected in a purely Minkowskian background. The
properties of wormholes are also sensitive to the signs of $r_g$ and $\Lambda_B$. In fact, as pointed out above, traversable wormholes (without event horizon and regular everywhere) exist as long as the number of charges satisfies the condition $N_q\lesssim 16.55(1-\delta_3)^2/(1+\delta_4^2)(1-\delta_3-\delta_4^2/2)$.  Therefore, the signs of $g$ and $\Lambda_B$ are likely to play a non-trivial role in the thermodynamics of these solutions when the magnitude of $r_g$ and $\Lambda_B$ can be freely specified. This aspect will be explored elsewhere.

\section{Conclusions}

In this work we have considered a quadratic gravity model formulated \`{a} la Palatini and coupled to a particular nonlinear theory of electrodynamics. While the former is motivated by  the existence of higher-order curvature corrections in different approaches to quantum gravity,
the latter is known to implement a confinement mechanism in flat space. The combination of these two models, for solutions with electric charge, leads to two main physically interesting results: i) appearance of a dynamically generated cosmological constant term at large distances and ii) existence of a set of solutions for which the curvature invariants are smooth everywhere and the space-time possesses a wormhole structure, representing an object that can be naturally interpreted in terms of Wheeler's geon \cite{Wheeler, lor13}. Thus, these nonsingular solutions represent wormholes with de Sitter asymptotics. Remarkably, our solutions (with or without $\Lambda_B$) admit an exact analytical expression and arise naturally from the nonlinear electromagnetic field rather than requiring exotic matter to generate  a pre-designed wormhole geometry (Morris-Thorne approach).

It should be stressed that as the surface $x=1$ is approached, the electromagnetic field grows well beyond the threshold of pair production. Thus, a more satisfactory description of the electrovacuum solutions and the associated geometry necessarily must take into account these quantum vacuum effects, which amounts to an ultraviolet modification of the matter Lagrangian. This point has already been worked out in detail in \cite{or13b} through a Born-Infeld electromagnetic Lagrangian coupled to the theory (\ref{eq:gravitytheory}), finding that the wormhole solutions are robust and in excellent qualitative agreement with the results found here \cite{or12a} and in \cite{lor13}. This paper, therefore, supports the consistency of the wormhole solutions when infrared and ultraviolet corrections are taken into account in the matter sector.

\section*{Acknowledgements}

G. J. O. is finantially supported by the Spanish grant FIS2011-29813-C02-02  and the JAE-doc program of the Spanish Research Council (CSIC). D. R. -G. is supported by CNPq (Brazilian agency) through grant 561069/2010-7 and acknowledges the hospitality and partial support of the theoretical physics group at Valencia U.

\end{document}